\def\cm{{\rm\thinspace cm}}

\def\erg{{\rm\thinspace erg}}

\def\keV{{\rm\thinspace keV}}

\def\Msun{\hbox{$\rm\thinspace M_{\odot}$}}

\def\s{{\rm\thinspace s}}

\def\ergpcmps{\hbox{$\erg\cm^{-3}\s^{-1}\,$}}

\def\spose#1{\hbox to 0pt{#1\hss}}
\def\approxlt{\mathrel{\spose{\lower 3pt\hbox{$\sim$}}
        \raise 2.0pt\hbox{$<$}}}
\def\approxgt{\mathrel{\spose{\lower 3pt\hbox{$\sim$}}
        \raise 2.0pt\hbox{$>$}}}

\documentstyle[psfig]{mn}

\title
[Cyclo--synchrotron emission from active
regions] {Cyclo--synchrotron emission from
magnetically dominated active regions above accretion discs}

\author[T.~Di~Matteo A.~Celotti and A.~C.~Fabian]
{T.~Di~Matteo$^1$,  A.~Celotti$^2$ and A.~C.~Fabian$^1$\\ 
{$^1$Institute of Astronomy, Madingley Road, Cambridge, CB3 OHA}\\
 $^2$ S.I.S.S.A., via Beirut 2--4, 34014 Trieste, Italy\\}

\begin{document}

\maketitle

\begin{abstract} We discuss the role of thermal cyclo--synchrotron
emission in a magnetically dominated corona above an accretion disk in
an active galaxy or a galactic black hole candidate. The dissipation
occurs in localized active regions around the central black hole.
Cyclo--synchrotron radiation is found to be an important process. In
the case of galactic black hole candidates, its emission can dominate
the inverse Compton scattering of the soft photon field produced by
the disc by thermal electrons. We discuss observational predictions
and the detectability of cyclo-synchrotron radiation both for these
sources and radio-quiet active galactic nuclei.
\end{abstract}


\section{Introduction}

Accretion onto a massive black hole is believed to provide the primary
energy source in active galaxies (AGN) and in some galactic X-ray
binaries (galactic black hole candidates, GBHC). It is still unclear
which fraction of the energy is released radiatively in an accretion
disk around the black hole compared to that dissipated in a tenuous hot
plasma (corona) above it.  Such dissipation probably occurs through
release of magnetic energy expelled from the disk by buoyancy effects.

Strong support for this picture comes from the interpretation of high
energy emission (from UV to $\gamma$--rays). Both the spectral
characteristics, including reflection components, and the energetics
are better explained within the context of disk-corona models (Haardt
\& Maraschi 1993). These assume the presence of a corona, the hot
thermal electrons of which Compton scatter the soft photon field
produced by, and self-consistently reprocessed in, the cold underlying
disk.  The observations of the ratios of UV and X-ray luminosities in
different objects require that the hot plasma covering factor is less
than unity, suggesting that the dissipation is localized in discrete
emitting regions (Galeev, Rosner \& Vaiana 1979; Haardt, Maraschi \&
Ghisellini 1994; Stern et al. 1995).

Within this framework we assume that the disk dissipates internally a
fraction $(1-f)$ of the accretion power, $\eta L_{\rm Edd}$, while the
remaining fraction $f$ is stored in magnetic field structures which
generate localized active regions. Given that here the energy is
transferred via the magnetic field, we expect it to be at least in
equipartition with the local radiation energy density. 
The magnetic field in such regions is given by
$B^2/8\pi={f\eta L_{\rm Edd}}/{N 4\pi (r_{\rm blob}R_{\rm s})^2 c}$,
where $N$ is the number of active regions at any time, $L_{\rm Edd}$
is the Eddington luminosity and $r_{\rm blob}$ is their typical size
in units of the Schwarzschild radius, $R_{\rm s}$. This corresponds to
a field of the order of

\begin{equation} B=3\times10^5\left(\frac{f\eta}{N}\right)^{1/2}
r_{\rm blob}^{-1}\left(\frac{M}{10^6\Msun}\right)^{-1/2}\;\; \;\;{\rm
G}, \end{equation} where $M$ is the black hole mass.  Energetic
electrons embedded in such a field radiate by cyclo-synchrotron
(hereafter CS) emission. As we show in the following sections, under
the temperature and density conditions required by the disk-corona
models, the CS radiation is partially self-absorbed. However, this
process can still significantly contribute to the cooling of the hot
plasma. Indeed, CS emission has been shown to be relevant in the case
of some GBHC (Cyg X-1 and GX339-4) and to explain the observed
rapid optical variability in GX339-4 (Fabian et al. 1982; Apparao
1984).

Self--absorbed CS emission has also been previously considered in the
context of accretion discs (Ipser \& Price 1982, Zdziarski 1986 and
references therein), while its contribution has often been neglected in
the context of localized magnetically dominated active
regions in a corona (e.g. Haardt et al. 1994). In this paper we
derive a useful diagnostic in this specific context.

In particular, 
we examine in some detail the role of CS radiation losses in
terms of both the electron cooling and the expected radiative
signatures in AGN and GBHC. In section 2 we estimate the CS cooling
timescale and compare it to the inverse Compton one. The emitted
spectrum and the possibility of detecting the CS radiation are
discussed in section 3. Section 4 summarizes our conclusions.

\section{Cyclo-synchrotron cooling}

In a thermal plasma, optically--thin synchrotron emission rises steeply
with decreasing frequency. Under most circumstances the emission becomes
self-absorbed and gives rise to a black-body spectrum below a critical
frequency $\nu_{\rm c}$. Above this frequency it decays exponentially as
expected from a thermal plasma, due to the superposition of cyclotron
harmonics.  The total cooling rate can be therefore estimated by taking
into account the blackbody emission up to a frequency $\nu_{\rm c}$ and
the CS one above $\nu_{\rm c}$.  We show here that the inclusion of this
latter component is important and can even exceed the emission from the
optically thick part.

In a plasma with electron density $n$ and dimensionless electron
temperature $\theta=k T/m_{\rm e}c^{2}$, embedded in a magnetic field
of strength $B$, the CS emissivity in the optically thin limit is given by
(e.g.  Pacholczyk 1970; Takahara \& Tsuruta 1982)

\begin{equation}
\epsilon_{\rm \nu}=5.57\times10^{-29}\;\frac{n\nu
I'(x)}{K_2(1/\theta)}\;\;\ergpcmps {\rm Hz^{-1}},
\end{equation}
where $x\equiv {2\nu}/{3\nu_0\theta^2}$, $\nu_0={eB}/{2\pi m_{\rm e} c}$, $K_2(1/\theta)$ is the modified Bessel
function  and  $I'(x)$ is expressed as the average over all
angles (with respect to the direction of $B$) for a mildly relativistic
plasma with an isotropic velocity distribution. It can be approximated as
(Mahadevan et al. 1996)  
\begin{equation} 
I'(x)=\frac{4.050}{x^{1/6}}\left(1+\frac{0.40}{{x}^{1/4}}+\frac{0.532}{x^{1/2}}\right) {\rm exp}(-1.889 x^{1/3}). 
\end{equation} 
The self-absorption frequency $\nu_{\rm c}$ is given by 
\begin{equation}
\nu_{\rm c}= \frac{3}{2}\nu_0\theta^2 x_{\rm m}, \end{equation}
 where $x_{\rm m}(\tau/B, \theta)\equiv x(\nu_{\rm c})$, can be
determined by equating the CS emission (eqn. 2) to the Rayleigh-Jeans
black-body emission from the surface of a sphere of radius $r_{\rm
blob}R_{\rm s}$, namely
\begin{equation}
\epsilon_{\nu}\frac{4\pi}{3}(r_{\rm blob}R_{\rm s})^3 =8\pi^2 m_{\rm
e}\nu^2\theta (r_{\rm blob}R_{\rm s})^2. 
\end{equation} 
The cooling rate for the optically thin CS component, $\Lambda_{\rm
thin}$, is determined by integrating $\epsilon_{\nu}$ over frequencies
above $\nu_{\rm c}$, while in the optically thick limit the cooling rate
can be approximated by $\Lambda_{\rm thick}\approx ({2\pi}/{3})m_{\rm
e} \theta {\nu_{\rm c}^3}/r_{\rm blob}R_{\rm s}$.
Fig.~1 shows the ratio of the two cooling rates
($\Lambda_{\rm thin}/\Lambda_{\rm thick}$) for different values of the
plasma temperature $\theta$ and the Thomson optical depth $\tau =
\sigma_{\rm T} n r_{\rm blob}R_{\rm s}$. As Fig. 1 illustrates, $\Lambda_{\rm
thin}$ is comparable to $\Lambda_{\rm thick}$ and for $\theta \approxgt
0.7$ it becomes the dominant cooling component. It
therefore needs to be taken into account when calculating the cooling
timescales for CS emission.
\begin{figure}
\centerline{\psfig{figure=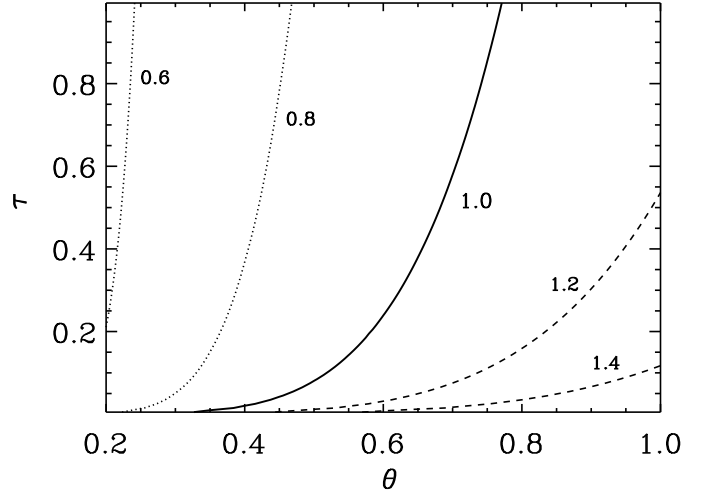 ,width=0.50\textwidth,angle=90}}
\caption{The  plot represents the ratio  $\Lambda_{\rm thin}/
 \Lambda_{\rm thick}$ as a function of $\theta$ and $\tau$ for CS emission for a typical GBHC. Curves are
labelled with the respective values of this ratio.}
\end{figure}

\subsection{Cooling timescales}

In order to establish the importance of typical CS losses in an active region,
here we estimate the timescales for CS and Compton emission.  The CS
cooling timescale, $t_{\rm CS}\simeq {n m_{\rm e} c^2
\theta}/({\Lambda_{\rm thick}+\Lambda_{\rm thin}})$ , is given by

\begin{equation}
t_{\rm CS}=\frac{3c^2}{2\pi\sigma_{\rm T}}\frac{\tau}{\Delta 
\nu_{\rm c}^3}.
\end{equation}
where $\Delta= \Lambda_{\rm thick}+\Lambda_{\rm thin}$, the total
cooling rate.  By substituting $\nu_{\rm c}$, we obtain

\begin{equation}
t_{\rm CS} \simeq
\frac{8.7\times10^{24}\tau}{\Delta B^3\theta^6x_{\rm
m}^3} 
\simeq \frac{0.3}{\Delta}\left(\frac{NM}{\eta f}\right)^{3/2}\frac{r_{\rm
blob}^3 \tau}{\theta^6x_{\rm m}^3} \;\s.
\end{equation}
$x_{\rm m} (\tau/B,\theta)$ is obtained by solving eqn.~(5) and is plotted as a
function of $\tau/B$ in Fig.~2.  

$t_{\rm CS}$ can be compared with the timescale for inverse Compton
cooling on the  radiation field of the disk-photons

\begin{equation}
t_{\rm iC}=\frac{m_{\rm e}c}{4\sigma_{\rm T}}\frac{1}{U_{\rm
rad}}=2.7\times10^{-9}\frac{r_{\rm cor}^2M}{\eta (1-f)} \;\;\;\s,
\end{equation}
where $U_{\rm rad}={(1-f)\eta L_{\rm Edd}}/{4\pi (r_{\rm cor} R_{\rm
s})^2 c}$ and $r_{\rm cor}$ is the total  extent of the disc emission in
units of $R_{\rm s}$.

\begin{figure}
\centerline{\psfig{figure=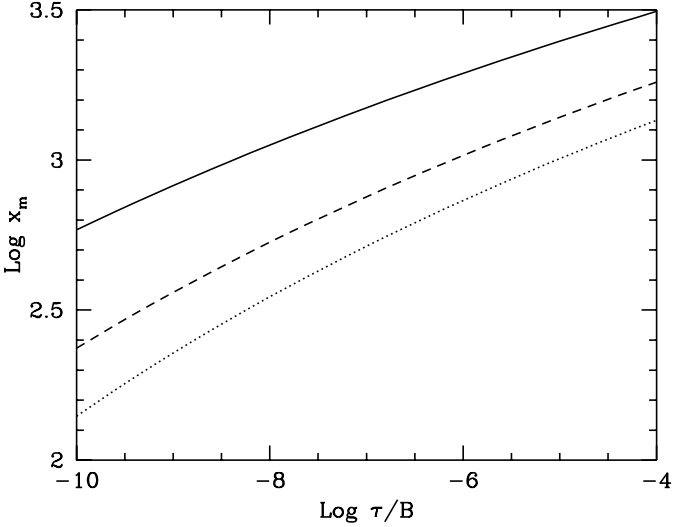 ,width=0.5\textwidth}}
\caption{$x_{\rm m}$, which determines the 
self--absorption frequency is shown versus $\tau/B$ for three values of
the electron temperature, namely $\theta=0.6$ -- solid line
$\theta=0.4$ -- dashed line and $\theta=0.2$ -- dotted line.}
\end{figure}

In order to compare these timescales we choose appropriate ranges for
the parameters $\tau$ and $\theta$. As mentioned above, according to
the `standard' models for high energy emission, X-rays are due to
Compton scattering of the soft photons by mildly--relativistic thermal
electrons. For Seyfert galaxies, the canonical (intrinsic) X--ray
spectral slope observed in AGN is of the order of $\alpha\simeq 1$,
when the effect of reflection onto the thick accretion disk is taken
into account. This spectral index and the typical energies of the high
energy ($\gamma$--ray) cut--offs (e.g. Gondek et al. 1996) suggest a
scattering plasma with $\tau\approxlt
0.5$. However, recent results (Poutanen et al. 1997, Zdziarski et al. 1997)
derive values of $\tau \approx 1$ and electron temperatures $\approx
100 \keV$.  The quantitative relation between $\theta$ and $\tau$ and
the energy spectral index $\alpha$, for optically--thin plasmas, is
given by $\alpha\simeq{-{\rm ln}P}/{{\rm ln}(1+4\theta+16\theta^2)}$
where $P$ is the average scattering probability $P=1 + {{\rm
exp}(-\tau)}/{2}\left({\tau}^{-1}-1\right) - {2\tau}^{-1} +
({\tau}/{2})E_1(\tau)$, and $E_1$ the exponential integral (Zdziarski
et al.  1994). This scattering probability is calculated strictly for
a slab geometry; we consider such an approximation acceptable, given
the uncertainties in the geometry of active regions and of dissipation
through reconnection events (which could possibly take place in a
sheet/shell geometry). By imposing $\alpha\sim 1$, we can therefore
select the region of parameter space for $\tau$ and $\theta$, shown in
Fig.~3, for which to estimate the cooling rates.

\begin{figure} \centerline{\psfig{figure=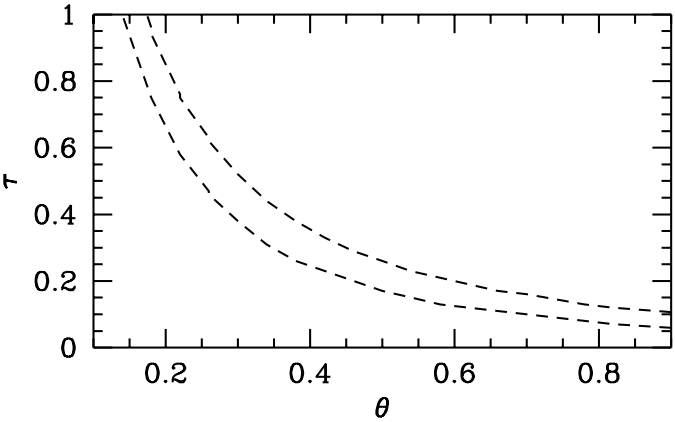 ,width=0.5\textwidth}}
\caption{The area between the two dashed curves represents the region of parameter space for
which combinations of $\theta$ and $\tau$ give $\alpha\simeq 1 \pm 0.1
$.}
\end{figure}

The ratio in the case of GBHC (we take $M=5\Msun$) and for
$\theta=0.4$ and
$\tau=0.3$ is given by 

\begin{eqnarray}
\frac{t_{\rm CS}}{t_{\rm iC}}&=&1.2\times 10^8 
\frac{1-f}{f^{3/2}\eta^{1/2}} N^{3/2}
\left(\frac{r_{\rm blob}}{r_{\rm cor}}\right)^2
\frac{r_{\rm blob}\tau}{\Delta \theta^6 x_{\rm m}^3}M^{1/2}\qquad \quad 
\nonumber \\
&\simeq&{0.5}\left(\frac{2}{\Delta}\right)
\left(\frac{1-f}{0.5}\right)\left(\frac{0.5}{f}\right)^{3/2}
\left(\frac{1}{\eta}\right)^{1/2}\nonumber
\qquad \\
&\times& \left(\frac{N}{10}\right)^{3/2}\left(\frac{r_{\rm
blob}}{1}\right)^3
\left(\frac{50}{r_{\rm cor}}\right)^{2}\left(\frac{\tau}{0.3}\right) \nonumber
\qquad \\
&\times&\left(\frac{0.4}{\theta}\right)^6\left(\frac{510}{{\rm x_m}}\right)^3
\left(\frac{M}{5 M_{\odot}}\right)^{1/2}.
\end{eqnarray}
It is worth stressing that, given the observational dispersion in
$\tau$ and $\theta$, for the above calculation we have chosen
intermediate values for these parameters (see Fig. 3). For sources
which extend in the range of higher temperatures ($\theta\approx 0.7$)
and lower values of the optical depth ($\tau=0.1$) the ratio becomes
of the order of $10^{-2}$.  In this same case, but for AGN with
$M=10^{6} \Msun$, it is given by ${t_{\rm CS}}/{t_{\rm iC}}\simeq 1$
where $x_{\rm m}=626$ and $\Delta=1.4$. This means that CS emission is
very likely to be relevant in GBHC and may be important for AGN,
depending on the values of the parameters.

\subsection{Physical considerations}
To estimate the size of the active regions we adopt the model
developed by Galeev et al. (1979) and recently re--considered by
Haardt et al. (1994) (see also Stern et al. 1995). There $r_{\rm blob}
\sim z_0/\alpha_{\rm v}^{1/3}$, where from the geometrically thin
accretion disk model (Shakura \& Sunyaev 1973), the disk scale height
$z_0$ is $\sim 9 L/L_{\rm Edd}S(r)R_{\rm S}$, (where
$S(r)=1-(3/r)^{1/2}$) and $\alpha_{\rm v}$ is the viscosity
parameter. In accordance with the Haardt et al. (1994) model, the
total number of active regions at any time is chosen to be $N\sim 10$.
This is consistent with the need to explain the amplitude of X-ray
fluctuations on
short timescales typically observed in AGN and GBHC.

Note that the importance of the CS emission depends strongly on the
relative size of the active regions as compared to the disk extension
$(r_{\rm blob}/r_{\rm cor})^2$. This ratio, however, is constrained by
the fraction of soft photons intercepted and Compton scattered in
these active regions (i.e. $N r^2_{\rm blob}$ cannot exceed $r^2_{\rm
cor}$).  Actually this covering factor is estimated in some cases to
be 50 per cent in sources where the UV/soft X--ray spectral component
strongly dominates over the hard X--ray emission (however in most
cases a covering factor significantly smaller than 50 per cent can be
observationally inferred). The ratio of the two sizes has therefore to
be at most $\approxlt 0.3$ if $N\sim 10$.
 
Under the assumptions adopted in eqn.~(9) for the
relative size of the two regions and the dissipation efficiency, the
magnetic energy density (as calculated from eqn.~(1)) largely
exceeds both the radiation energy density, $U_{\rm rad}$, and the
thermal particle (electron/positron) one, $\sim n m_{\rm e} c^2
\theta$, which are of the same order. 
We note here that the powers released into
magnetic field and disk radiation have been assumed to be of the order of
$L_{\rm Edd}$. Sub-Eddington powers, i.e. $\eta\approxlt 1$, clearly
influence the relative importance of the CS emission. However, as
shown in eqn.~9, even for $\eta \sim$ 0.1, the ratio of the
cooling timescales varies only by a factor of 3.

The computation of the temporal evolution of the magnetic dissipation
and consequent particle acceleration is beyond the aim of this
paper. Here we can simply estimate the reconnection timescale for the
stored magnetic field energy to be converted into particle
energy. This can be written as $t_{\rm rec}\sim r_{\rm blob} R_{\rm
s}/v_{\rm A}$, where the reconnection velocity is taken of the order
of the Alfven speed $\sim c$ (Priest \& Forbes 1986).  Note that,
unlike other authors, here we are not assuming a priori that the whole
of the energy dissipated into the corona ($fL_{\rm Edd}$) is instantly
released as radiation: the accretion energy is stored as magnetic
energy in the active region and no efficiency is assumed in the
conversion of the latter into radiation.  It is then possible to
estimate such efficiency by comparing the total energy stored in the
magnetic field with the energy radiated.  Given that in our model the
cooling timescale of the plasma is of the order of the reconnection
timescale $t_{\rm rec}$, this corresponds to comparing $U_{\rm B}
(4\pi/3) (r_{\rm blob}R_{\rm S})^3$ and $ (r_{\rm blob}R_{\rm S}/c)
L_{\rm blob}$, where $L_{\rm blob}$ is the total luminosity of an
active region (all the radiative processes are taken into account).
By computing $L_{\rm blob}$, we obtain a magnetic energy $\sim 90 - 200$
$\times$ the radiated one, implying an extremely low efficiency in the
energy conversion.  The magnetic energy could then be efficiently
converted if multiple reconnection events within the same active
region occur.
 
To summarize this section, we find that CS emission from thermal
electrons can play an important role if the active regions, or blobs,
are localized and relatively small. Inverse Compton cooling is
favoured instead in the case where the difference between the `slab'
and `patchy' corona models becomes less extreme, either for a lower
dissipation fraction $f$, or for higher covering factor of the active
blobs.  A similar trend is obtained for sub-Eddington accretion.
 
\section{Cyclo--synchrotron emission}

We now consider the intensity of CS emission and its possible
detectability, by comparing the predictions of the model with the spectral
energy distributions of some AGN and GBHC. 

The CS spectra for different parameters $\tau$ and $\theta$ are shown
in Fig.~4a for a typical AGN. The plasma properties are again selected
to produce a Comptonized spectrum (of the CS photons) with $\alpha\sim
1$. For comparison, a blackbody spectrum with energy density
equivalent to the luminosity produced in the disk ($\eta (1-f)L_{\rm
Edd}$) is also reproduced in the figure. Fig.~4a
clearly shows that both $\nu_{\rm c}$ and the luminosity at this
peak frequency depend primarily on $\theta$ (when the condition
$\alpha\sim1$ is satisfied). In particular if $\theta$ decreases, the
luminosity follows it.

\begin{figure} \centerline{\psfig{figure=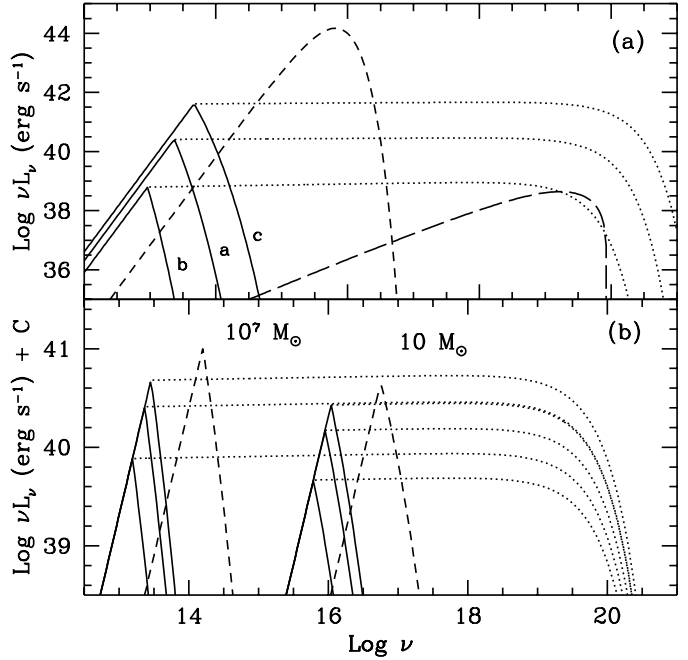 ,width=0.5\textwidth}}
\caption{(a) Dependence of the CS spectral distribution (in $\nu
L_{\nu}$) on the physical
parameters of the plasma, namely the temperature $\theta$ and optical
depth $\tau$. The spectra are estimated for a typical AGN, with $M=
10^7 M_{\odot}$, $f=0.5$ and $r_{\rm blob}=1$. The three solid curves
represent the CS emission for $\theta = 0.4$, $\tau=0.3$
(curve labelled a); $\theta=0.2$, $\tau=1$ (curve b); and $\theta =
0.74$, $\tau=0.1$ (curve c). For comparison, the scattered inverse
Compton (dotted lines) and bremsstrahlung (long-dash line) components
are also reported for case a. The short-dash curve represents instead
the blackbody emission from the disk. (b) The CS emission dependence
on the black hole mass, fraction $f$ of the accretion power magnetically
dissipated and the dimension of the active regions. As labelled, the
two sets of curves are computed for the cases $M=10^7 M_{\odot}$ and
$M=10 M_{\odot}$ (where in the latter one the vertical axis has been
displaced by a factor $10^4$). For each set the three curves are
computed for $f=0.2, 0.5, 0.8$, from bottom to top,
respectively. Finally, the dashed lines refer to $r_{\rm blob}=0.1$. In
all these cases $\theta=0.4$ and $\tau=0.3$.}
\end{figure}

The analogous plot in Fig.~4b shows the dependence on different central
masses, dimensions of the active regions and the dissipation fraction $f$. A
decrease in mass implies a lower luminosity, but due to the increase in
the equipartition magnetic field, leads to a higher self--absorption
frequency. Clearly more coronal dissipation (higher $f$) gives rise to an
enhancement in the frequency and luminosity of CS emission. The same
effects are produced by reducing the size of the emitting regions. 

\subsection{Observational predictions}

As discussed above, the peak of the CS emission is located in the
IR--EUV band, strongly dependent on the black hole mass. At the same
time the intensity of the emitted luminosity can be significant from
an observational point of view.  For this reason, we compare the
predicted spectra with the spectral energy distribution (SED) of a few
GBHC and 
radio-quiet AGN. The data (not simultaneous) have been
collected from the literature in order to cover the broader possible
spectral range. The CS and its relative Compton scattered component,
have been self--consistently computed taking into account the relative
importance of these processes with respect to the cooling by inverse
Compton on the soft photons from the disk, as estimated above
(see eqn.~9). Note
that in the case of GBHC the radiation energy density in CS photons
can exceed that of the disk.

\begin{figure} \centerline{\psfig{figure=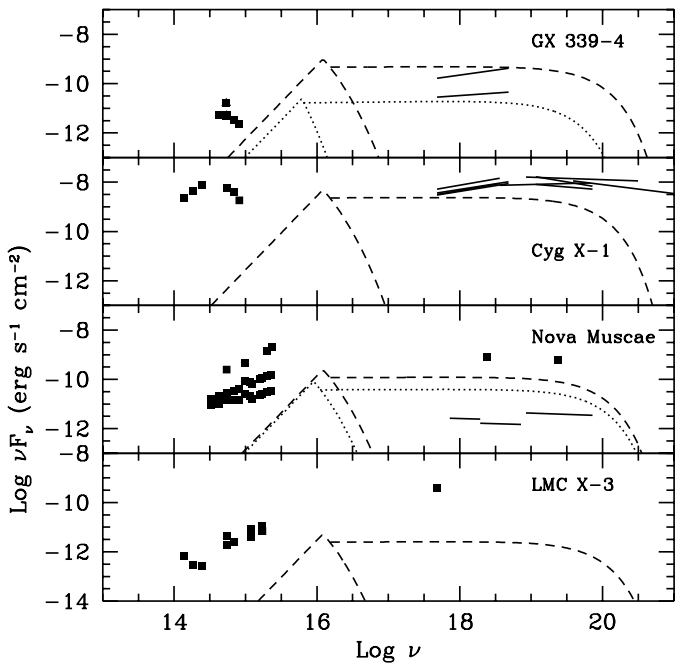 ,width=0.5\textwidth}}
\caption{Comparison of the CS and inverse Compton emission  
with the SED of four GBHC, namely GX 339-4, Cygnus X-1, Nova
Muscae and LMC X--3. The CS emission is estimated for a mass of $10
M_{\odot}$, $f=0.5$, $r_{\rm blob}=1$, $\theta=0.4$ and $\tau=0.3$--dashed
line. The dotted-line in GX~339-4 is calculated for $\theta=0.15$ and
$\tau=1$ whereas the dotted-line in Nova Muscae is for $r_{\rm
blob}=0.8$.  Data are from: Miyamoto et al.  (1992), Makishima et
al. (1986), Motch et al. (1983) (GX 339-4); Miyamoto et al. (1992),
Salotti et al. (1992), Barr, White
\& Page (1985), Ling et al. (1983), Treves et al. (1980) (Cygnus
X--1); King, Harrison \& McNamara (1996), Gilfanov et al. (1993),
Shrader \& Gonzalez--Riestra (1993), Cheng et al.  (1992) (Nova
Muscae); Treves et al. (1990), Treves et al.  (1988) (LMC X--3).}
\end{figure}
\begin{figure} \centerline{\psfig{figure=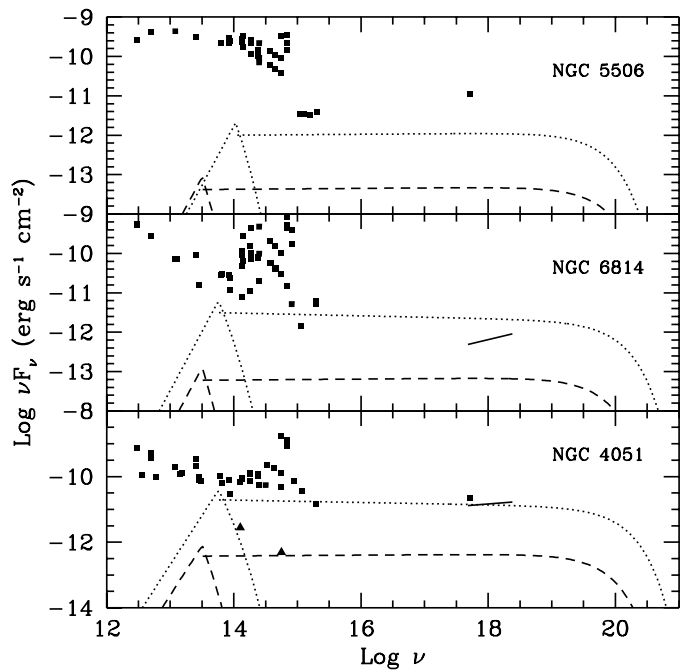,width=0.5\textwidth}}
\caption{Analogous to Fig.~5, for the three bright Seyfert 1 galaxies NGC
5506, NGC 6814 and NGC 4051.  The predicted emission is computed for
$\theta=0.4$, $\tau=0.3$, $f=0.8$, $r_{\rm blob}=0.8$ and $M=10^7
M_{\odot}$ --dashed lines. In NGC 6814 and NGC 4015 the dotted-line is
determined with $\theta=0.74$ and $\tau=0.1$. In NGC 5506 the dotted
curve is computed for a magnetic field 4$\times B$.  The two triangles in
the SED of NGC
4051
indicate the variability limits reported by Done et al. (1990) (see
the text). Data are from the following references: Kinney et al. (1993),
White \& Becker (1992), Fabbiano, Kim
\& Trinchieri (1992), De Vaucouleurs et al. (1991), Moshir et
al. (1990), Soifer et al. (1989), McAlary et al. (1983), Glass (1981),
Glass (1979) (NGC 5506); Reynolds (1997), Fabbiano, Kim \& Trinchieri
(1992), Celotti, Ghisellini \& Fabian (1991), Moshir et al. (1990),
McAlary \& Rieke (1988), McAlary et al.  (1983), Glass (1981), Glass
(1979), Rieke (1978), Penston et al. (1974) (NGC 6814); Reynolds
(1997), Fabbiano, Kim \& Trinchieri (1992), De Vaucouleurs et al.  (1991),
Soifer et al. (1991), Gregory \& Condon (1991), Moshir et al. (1990),
Done et al. (1990), De Vaucouleurs \& Longo (1988), Ward et
al. (1987), Ficarra et al. (1985), Balzano \& Weedman (1981), Lebofsky
\& Rieke (1979), McAlary et al. (1979), Rieke (1978), Rieke
\& Low (1972), Wisniewski \& Kleinmann (1968) (NGC 4051).}
\end{figure}

The results of the comparison are presented in Figs.~5,6 for GBHC
(GX~339--4, Cyg X--1, Nova Muscae and LMC X--3) and AGN (Seyfert 1
galaxies: NGC 5506, NGC 6814 and NGC 4051), respectively
\footnote{Note that the masses adopted differ from those used in
eqn.~(9).}.  In the case of GHBC, the peak of the CS emission is in
the EUV band, currently not observed, with the Compton--scattered CS
flux which in some cases could exceed the X--ray flux level. The high
energy tail of the CS emission can however be important at
$0.1\keV$. The CS emission itself can be comparable (within a factor
of $\sim$ 3 for GX~339--4) with the optical--UV measured fluxes and
possibly contribute to the soft X--ray excess. This implies that the
CS component could give rise to a detectable flux variability.
 Anti-correlation between optical and X-ray variations
observed in GX~339-4 (Motch et al. 1983) can be consistent with the
model, with a decrease in the field intensity after a reconnection
event.

In AGN, the CS spectral
component is expected to contribute to the IR--near IR emission, with
a flux which can be more than $\sim$ 10 per cent of the observed one.
Particularly interesting is the case of NGC 4051 for which
constant IR and optical fluxes (within 1 per cent) have been observed
simultaneously with a factor 2 fluctuation in the X-ray band (Done et
al.  1990).  In Fig.~6 we report these variable IR and optical flux
constraints, which indeed set strong upper limits to any low energy
radiation produced from the same population of electrons responsible
for the X--ray emission. The detection of the CS component can be then
a very powerful tool for investigating the physical conditions and the
role of magnetic field in the inner corona of these sources. In
particular, we stress that with an increase of only a factor of 4 in
the field intensity, the predicted flux (and frequency) can increase
to levels comparable with the data and the variability limits
(e.g. the case of NGC 5506, see Fig.~6).  

Note that here we used sets of parameters in the figures just
as indicative predictions of the model. In particular, it is important to
point out that even within the same source, the magnetic energy
density is most likely to differ between the active regions.

We stress that the CS spectral component is expected to give rise to strong
variability on short timescales as determined by the typical size of the
active region ($\sim R_{\rm S}/c$) \footnote{ Interestingly, some hints of
small scale fast variations have been observed as intraday optical
variability in several radio--loud sources and in some
radio--quiet ones (e.g. Dultzin--Hacyan et al. 1992; Sagar, Gopal--Krishna
\& Wiita 1996)},
provided that the number of active
regions $N$ is small. Furthermore, the variability is predicted to be
simultaneous to the X-ray one.  A high degree of linear polarization,
rapidly variable both in intensity and polarization angle, constitutes
a further signature of the presence of thermal CS radiation.

We also stress that it is possible that the different states of
individual GBHC (eg. Cyg X-1 and GX339-4) could be due to variations
in the ratio and dominance of inverse Compton scattering of CS and
external disk photons, e.g. to changes in $B$ or $r_{\rm blob}$.

\section{Summary}

We have examined the role of CS emission in active regions, which are
believed to be the loci of magnetic energy dissipation in the coronae of
accretion disks. The particles, possibly energized through 
reconnection, are expected to have a thermal distribution, as also
suggested by the observed high energy emission of radio-quiet AGN and GBHC. 

The CS process can be an important cooling mechanism if the magnetic
energy dissipation can occur in a much smaller volume than the soft
thermal photon field which is distributed over the whole inner
disk. Locally, the absorbed CS can then compete with Compton cooling
on the photons produced by the disk. As a consequence of this, the
effect of CS cooling on thermal electrons has to be considered in
Comptonization models.  We also note that self-absorbed CS
emission can act as a rapid and efficient thermalizing mechanism for
any initial electron distribution (Ghisellini, Guilbert \& Svensson 
1988).

Typical black hole masses and physical parameters of the thermal plasma
are sufficient to produce observable radiative signatures, in both
classes of sources, when $\sim$ 50 per cent of the accreting power is
dissipated
into the corona. Even in the most unfavourable case of AGN, CS variable
and polarized emission is predicted in the near IR band at a level
above a few to tens of per cent. In GBHC the spectrum would be
observable  in the EUV band. Future monitoring in
these bands with high temporal resolution, especially simultaneous with X-rays
observations, can therefore  provide crucial information
on the physical conditions and main dissipation mechanisms in the very
central regions of compact sources.

\section*{Acknowledgements}

This research has made use of the NASA/IPAC extragalactic database
(NED), which is operated by the Jet Propulsion Laboratory, California
Institute of Technology, under contract with the National Aeronautic
and Space Administration. We thank Dr. Juri Poutanen for useful
comments. For financial support, we acknowledge PPARC and Trinity
College, Cambridge (TDM), the Italian MURST (AC) and the Royal Society
(ACF).


\begin{thebibliography}{}
\bibitem{} Apparao K.M.V., 1984, A\&A, 139, 377
\bibitem{} Balzano V.A., Weedman D.W., 1981, ApJ, 243, 756
\bibitem{} Barr P., White N.E., Page C.G., 1985, MNRAS, 216, 65p
\bibitem{} Celotti A., Ghisellini G., Fabian A.C., 1991, MNRAS, 251, 529
\bibitem{} Cheng F.H., Horne K., Panagia N., Shrader C.R., Gilmozzi R.,
Paresce F., Lund N., 1992, ApJ, 397, 664
\bibitem{} De Vaucouleurs, A., Longo G., 1988, Catalogue of Visual and Infrared Photometry of Galaxies from 0.5 micron to 10 micron
\bibitem{} De Vaucouleurs G., De Vaucouleurs A., Corwin J. H.G., Buta R.J.,
 Paturel G., Fouque P.,  1991, Third Reference Catalogue of Bright Galaxies, Version 3.9
\bibitem{} Done C., Ward M.J., Fabian A.C., Kunieda H., Tsuruta S.
Lawrence A., Smith M.G., Wamsteker W., 1990, MNRAS, 243, 713
\bibitem{} Dultzin--Hacyan D., Schuster W.J., Parrao L.,
Pena J.H., Peniche R., Benitez E., Costero R., 1992, AJ, 103, 1769
\bibitem{} Fabbiano G., Kim D.W., Trinchieri T., 1992, ApJS, 80, 531
\bibitem{} Fabian A.C., Guilbert P.W., Motch C., Ricketts M., Ilovaisky
S.A., Chevalier C., 1982, A\&A, 111, l9 
\bibitem{} Ficarra A., Grueff G., Tomassetti T., 1985, A\&As, 59, 255
\bibitem{} Galeev A.A., Rosner R., Vaiana G.S., ApJ,  1979, 229, 318
\bibitem{} Ghisellini G., Guilbert P.W., Svensson R., 1988, ApJ, 335, L5
\bibitem{} Gilfanov M., et al., 1993, A\&ASS, 97, 303 
\bibitem{} Glass I.S., 1979, MNRAS, 186, 29p
\bibitem{} Glass I.S., 1981, MNRAS, 197, 1067
\bibitem{} Gondek D., Zdziarski A.A., Johnson W.N., George I.M., McNaron-Brown K., Magdziarz P., Smith D., Gruber D.E., 1996, MNRAS, 282, 646
\bibitem{} Gregory P.~C., Condon J.~J, 1991, ApjS, 75, 1011
%\bibitem{} Haardt F., Done C., Matt G., Fabian A.~C., 1993, ApJ, 411, L95
\bibitem{} Haardt F., Maraschi L., 1993, 413, 507
\bibitem{} Haardt F., Maraschi L., Ghisellini G., 1994, ApJ, 432, L95
\bibitem{} Ipser R.J., Price R.H., 1982, ApJ, 255, 654 
\bibitem{} King N.L., Harrison T.E., McNamara B.J., 1996, AJ, 111, 16
\bibitem{} Kinney A.L., Bohlin R.C., Calzetti D., Panagia N., Wyse R.F., 1993, ApJS, 86, 5
%\bibitem{} Kleinmann D.~E. \& Low F.~J., 1970a, ApJ, 159, L165
%\bibitem{} Kleinmann D.~E. \& Low F.~J., 1970b, ApJ, 161, L203
\bibitem{} Lebofsky M.J., Rieke G.H., 1979, ApJ, 229, 111
\bibitem{} Ling J.C., Mahoney W.A., Wheaton W.A., Jacobson A.S.,
Kaluzienski L., 1983, ApJ, 275, 307
\bibitem{} Mahadevan R., Narayan R., Yi I., 1996, ApJ, 465, 327
\bibitem{} Makishima K., Maejima Y., Mitsuda K., Bradt H.V., Remillard
R.A., Tuohy I.R., Hoshi R., Nakagawa M., 1086, ApJ, 308, 635
\bibitem{} McAlary C.W., Rieke G.H., 1988, ApJ, 333, 1
\bibitem{} McAlary, C.W., McLaren R.~A., Crabtree D.~R., 1979, ApJ, 234, 471
\bibitem{} McAlary C.W., McLaren R.A., McGonegal R.J., Maza J., 1983,
ApJS, 52, 341
\bibitem{} Mihamoto S., Kitamoto S., Sayuri I., Negoro H., Terada K.,
1992, ApJ, 391, L21
\bibitem{} Moshir M., et al., 1990, Infrared Astronomical Satellite
catalogs, The Faint Source Catalog, Version 2.0
\bibitem{} Motch C., Ricketts M.J., Page C.G., Ilovaisky S.A., Chevalier
C., 1983, A\&A, 119, 171
\bibitem{} Pacholczyk A.G., 1970, in Radio Astrophysics, Freeman (San 
Francisco) 
\bibitem{} Penston M.V., Penston M.J., Selmes R.A., Becklin E.E.,
Neugebauer G., 1974, MNRAS, 169, 357
\bibitem{} Poutanen J., Svensson R., Stern B., 1997,  Proceedings of 2nd INTEGRAL Workshop, San Malo, France, p.401
\bibitem{} Priest E.R., Forbes T.G., 1986, J. Geophys. Res., 91, 5579
\bibitem{} Reynolds C.S., 1997, MNRAS,  in press
\bibitem{} Rieke G.H., 1978, ApJ, 226, 550
\bibitem{} Rieke G.H., Low F.J., 1972, ApJ, 176, L95
\bibitem{} Sagar R., Gopal--Krishna, Wiita P.J., 1996, MNRAS, 281, 1267
\bibitem{} Salotti L., et al., 1992, A\&A, 253, 145
\bibitem{} Shakura N.I., Sunyaev R.A., 1973, A\&A, 24, 337
\bibitem{} Shrader C.R., Gonzalez--Riestra R., 1993, A\&A, 276, 373
\bibitem{} Soifer B.T.,  Boehmer L., Neugebauer G., Sanders D.B., 1989,
AJ, 98 ,766 
\bibitem{} Stern B.E., Poutanen J., Svensson R., Sikora M., Begelman
M.C., 1995, ApJ, 449, L13
\bibitem{} Takahara F., Tsuruta S., 1982, Prog. Theoret. Phys., 67, 485
\bibitem{} Treves A., Belloni T., Bouchet P., Chiappetti L., Falomo R.,
Maraschi L., Tanzi E.G., 1988, ApJ, 335, 142
\bibitem{} Treves A., et al., 1980, ApJ, 242, 1114
\bibitem{} Treves A., et al., 1990, ApJ, 364, 266
\bibitem{} Ward M., Elvis M., Fabbiano G., Carleton N.P., Willner S.P.,
Lawrence A., 1987, ApJ, 315, 74
\bibitem{} White R.L., Becker R.H., 1992, ApJS, 98, 766
\bibitem{} Wisniewski, W.Z., Kleinmann, D.E., 1968, AJ, 73, 866
\bibitem{} Zdziarski A.A, 1986, ApJ, 289, 514
\bibitem{} Zdziarski A.A., Fabian A.C., Nandra K., Celotti A., Rees
M.J., Done C., Coppi P.S., Madejski G.M., 1994, MNRAS, 269, L55
\bibitem{} Zdziarski A.A., Johnson W.N., Poutanen J., Magdziarz P., Gielrinski M., 1997, Proceedings of 2nd INTEGRAL Workshop, San Malo, France, p. 373
\end{thebibliography}
\end{document}